\documentclass[%
 reprint,
superscriptaddress,
nofootinbib,
 amsmath,amssymb,
 aps,
]{revtex4-1}

\usepackage[pdftex,colorlinks]{hyperref}
\usepackage[dvipsnames]{xcolor}
\definecolor{phthaloblue}{rgb}{0.0, 0.06, 0.54}
\hypersetup{
    colorlinks=true,
    linkcolor=NavyBlue,
    citecolor=NavyBlue,
    filecolor=NavyBlue,
    urlcolor=NavyBlue,
    }
\usepackage[pdftex]{graphicx}
\usepackage{bm} 
\usepackage{cancel}
\usepackage{here}
\usepackage{comment,braket}
\usepackage{epsf}
\usepackage{amsmath}
\usepackage{graphics} 
\usepackage{subfigure}
\usepackage{amsfonts}
\usepackage{amssymb}
\usepackage{latexsym}
\usepackage{color}
\usepackage{natbib}
\usepackage[normalem]{ulem}
\usepackage{orcidlink}
\usepackage{cleveref}
\usepackage{physics}
\usepackage{upgreek}

\newcommand{\ba}{\begin{array}}
\newcommand{\ea}{\end{array}}

\newcommand{\ipc}[1]{\textcolor{NavyBlue}{#1}}

\begin{document}


\title{Hidden symmetries and the universality of Hawking radiation}

\author{Ivan Pérez-Castro\,\orcidlink{0009-0003-1060-6858}}
\email[~]{ivan.perez.c@cinvestav.mx}
\affiliation{Departamento de F\'{\i}sica, Centro de Investigaci\'on y de Estudios Avanzados del Instituto Politécnico Nacional \\ Apartado Postal 14-740, 07000, Ciudad de M\'exico, M\'exico.}

\author{Abdel Pérez-Lorenzana\,\orcidlink{0000-0001-9442-3538}}
\email[~]{abdel.perez@cinvestav.mx}
\affiliation{Departamento de F\'{\i}sica, Centro de Investigaci\'on y de Estudios Avanzados del Instituto Politécnico Nacional \\ Apartado Postal 14-740, 07000, Ciudad de M\'exico, M\'exico.}


\begin{abstract}
Quantum tunneling of particles across a black hole (BH) horizon provides an intuitive interpretation of Hawking radiation (HR), yielding a conceptually simple derivation of the thermal spectrum within a semiclassical framework. However, for BHs obeying the no-hair theorem, usual implementations of the complex-path (Hamilton–Jacobi) method typically rely on approximating the metric near the BH horizon. In this work, we demonstrate that the Hawking spectrum can be derived without such approximations by exploiting gauge invariance in quantum field theory and the separability of the Hamilton–Jacobi (HJ) equation arising from hidden symmetries of these spacetimes. This perspective shows explicitly why the BH behaves as an effective $(1+1)$-dimensional blackbody at the event horizon, provides the conditions needed to extend the HR to higher-spin particles, and offers a framework that can naturally incorporate back-reaction and quantum-gravity effects.
\end{abstract}

\maketitle


\section{Introduction}\label{sec:intro}

The quantum phenomenon known as Hawking radiation (HR), arising in extreme gravitational environments such as black holes (BHs), stands as one of the most remarkable results in theoretical physics. Since Hawking’s original derivations~\cite{Hawking1974black, Hawking1975particle}, a variety of approaches have been developed to reproduce the Hawking spectrum. From the Euclidean quantum gravity derivation~\cite{Gibbons1977action}, HR can be interpreted as a pair-creation process near the event horizon where one particle tunnels through classically forbidden trajectories. In this setting, two semiclassical frameworks have been developed: the null geodesic method~\cite{Parikh2000hawking, Parikh2004secret, Zhang2005hawking, Zhang2006charged, Akhmedov2008subtleties, Umetsu2010tunneling, Vanzo2011tunneling} and the complex path approach~\cite{Srinivasan1999particle, Shankaranarayanan2001method}, which uses the Wentzel–Kramers–Brillouin (WKB) approximation to derive the HJ equation~\cite{Angheben2005hawking, Akhmedov2006hawking, Kerner2008fermions, Banerjee2008quantum}. An alternative framework is provided by the anomaly cancellation method~\cite{Robinson2005relationship, Iso2006hawking}.

Cancellation of gauge and gravitational anomalies provides a compelling derivation of HR by enforcing gauge invariance and general coordinate covariance at the quantum level~\cite{Iso2006hawking, Iso2006anomalies}. Although ambiguities in this approach related to a constant of integration have been addressed through suitable modifications~\cite{Morita2009modification}, important challenges remain. In particular, it is challenging to incorporate possible low-energy effects, such as violations of Lorentz invariance or other quantum gravity effects. By contrast, these effects can be incorporated within the complex-path method (see, for example, ~\cite{ Chen2013fermion, Liu2014fermions, Chen2014effects, Yang2016lorentz, Li2016massive, Priyobarta2022modified, Singh2024maxwell, P2025modified}).

In the complex-path method, the analysis of rotating BHs typically relies on fixing the polar angle near the event horizon at the metric level, as inherited from the null-geodesic approach, where the integral expression develops singularities associated with the ergosphere, thereby requiring a near-horizon approximation~\cite{Kerner2006tunnelling}. However, in the case of rotating BH solutions, such as Kerr~\cite{Kerr1963gravitational} and Kerr–Newman (KN)~\cite{Newman1965metric}, the separability of the HJ equation reflects a fundamental property of the spacetime geometry and is manifest in Boyer–Lindquist coordinates~\cite{Carter1968global, Carter2009republication}.
From this perspective, it is somewhat counterintuitive that a near-horizon approximation of the HJ solution is still required to derive HR. Motivated by this observation, in this paper we show that, by exploiting gauge invariance of quantum fields, we can establish well-defined wave equations and then derive the Hawking spectrum without relying on such approximations.

The rest of the manuscript is organized as follows. In Sec.~\ref{sec:Complex-path-HR}, we review the complex-path method used to determine the Hawking spectrum. In Sec.~\ref{sec: quantum-fields}, we discuss gauge invariance of known physical quantum fields in flat spacetime and outline how additional fields may arise in supersymmetric scenarios and in higher-spin theories. In Sec.~\ref{sec: quantum-fields-curved}, we present the treatment of quantum fields in curved spacetime. In Sec.~\ref{sec:HJ and HR}, by exploiting the hidden symmetries responsible for the separability of the HJ equations, we show how HR emerges. In Sec.~\ref{sec: HR for SM}, we provide a direct derivation of the Hawking spectrum for the particle content of the Standard Model (SM). We then extend the discussion in Sec.~\ref{sec: HR BSM} beyond the \ipc{SM} scenarios, emphasizing the challenges in establishing a universal description of HR for higher-spin fields. 
In Sec.~\ref{sec: HR-AC}, we present a discussion of the physical implications and possible extensions of the results.
Finally, our conclusions are presented in Sec.~\ref{sec: Conclusion}.

\section{Complex path method for Hawking radiation} \label{sec:Complex-path-HR}
In this section, we present the key elements for establishing the complex path method~\footnote{Sometimes referred to as the HJ method.} to derive HR.

In early formulations of quantum tunneling through the event horizon of a BH, several conceptual difficulties arise even for the Schwarzschild BH:
\begin{enumerate}
    \item If the pair is created behind the horizon, neither of the particles can tunnel through the horizon, because the tunneling process in quantum mechanics is described via the solution of a Cauchy problem~\cite{Srinivasan1999particle} and has to be causal, while passing through the horizon is acausal~\cite{Akhmedov2008subtleties}.
    \item If the pair is created outside a horizon, the time for one of the particles to cross the event horizon is infinite for the stationary distant observer. However, this same observer should see the radiation from the BH in a finite time after the collapse.
\end{enumerate}
These subtleties were addressed in~\cite{Parikh2000hawking} by incorporating energy conservation and allowing for a dynamical background. In this picture, the emission process is associated with a shrinking of the event horizon, so that the created particle effectively materializes just outside the event horizon and can propagate to infinity.

Under these conditions, the initial radius of the event horizon is greater than the final one, $r_{\rm in} > r_{\rm out}$, since the BH loses mass during the emission of the s-wave. Consequently, the radius of the event horizon decreases. We parameterize the trajectory across the horizon as follows,
    \begin{equation}
        r_{\rm in} = r_H(M) - \epsilon \quad \text{and} \quad r_{\rm out} = r_H(M-\omega) + \epsilon \, . \label{radii}
    \end{equation}
where $r_H(M)$ denotes the location of the  BH event horizon before emission, and $\epsilon$ is an infinitesimal positive parameter.

Within this framework, deriving HR has been argued to require multiple pair-creation events rather than a single process~\cite{Akhmedov2006hawking}; for a virtual pair created just inside the horizon, the positive-energy particle must tunnel outward across the horizon, while the ``negative-energy'' partner propagates inward, reducing the mass of the BH. Conversely, for a virtual pair created just outside the horizon, it is the negative-energy particle that must tunnel inward through the horizon, while the positive-energy particle escapes to infinity. Then, \emph{contrary to classical intuition, the horizon represents a two-way barrier when considering virtual particle pairs}~\cite{Akhmedov2006hawking}. This would imply that the probability of tunneling inward, $P[in] \neq 1$ corresponds to one pair, and the probability of tunneling forward, $P[out] \neq 1$ corresponds to distinct tunneling processes, each associated with a different virtual pair.

This issue can be resolved using the standard semiclassical treatment of particle production in BH spacetimes developed by Hartle and Hawking~\cite{Hartle1976path}. In this approach, the Schwarzschild time coordinate $t$ is analytically continued to the complex plane, relating the Lorentzian spacetime to its Euclidean section. Requiring regularity of the Euclidean manifold enforces periodicity in imaginary time, which yields a thermal spectrum with a temperature identified as the Hawking temperature. This regularity condition also selects positive-frequency (positive-energy) modes, thereby fixing the physical vacuum state.

The analytic continuation of the time coordinate implies that the probability of particle emission from the past horizon, $P[out]$, differs from the probability of absorption into the future horizon, $P[in]$. Their ratio takes the form
    \begin{equation}
        \Gamma = \frac{P[out]}{P[in]} = e^{-\beta \omega}
    \end{equation}
where $\omega$ is the energy of the \ipc{emitted} particles and $\beta$ is the standard inverse Hawking temperature (with $k_B =1$)~\cite{Srinivasan1999particle}. This relation is naturally interpreted as a Boltzmann factor, indicating that the emitted radiation follows a thermal spectrum, analogous to that of a system in equilibrium with blackbody radiation. Then, we always take $P[in]=1$ for the incoming mode; the consistency of this interpretation has been discussed extensively in the literature; see, for instance,~\cite{Vanzo2011tunneling}.

Due to the infinite blueshift near the horizon, the wavelength of any wavepacket becomes arbitrarily small, making the geometrical optics approximation highly reliable. This allows us to work directly in a particle picture, avoiding the need for the full Bogoliubov formalism. In the semiclassical limit, the WKB approximation applies, relating the tunneling amplitude to the imaginary part of the action at the stationary phase. The emission rate, $\Gamma$, is then given by the square of the tunneling amplitude:
    \begin{equation}
        \Gamma = e^{-2{\rm Im} I} \approx e^{-\beta \omega} \, .
    \end{equation}
Early studies argued that the action $I$ could be expanded to account for quantum corrections~\cite{Banerjee2008quantum, Majhi2009fermion}. Nevertheless, as noted in~\cite{Wang2010hawking, Chatterjee2009hawking, Yale2011exact, Yale2011there}, the proper definition of energy within the HJ framework, $\omega = - \partial_t I$ shows that quantum corrections do not originate from the expansion of $I$~\cite{Vanzo2011tunn}.

Importantly, the incoming mode does not generally acquire an imaginary contribution. When such a contribution does appear, it typically indicates that the chosen coordinate system fails to provide a regular description across the horizon (see, e.g., Ref.~\cite{P2025modified}). 
In practice, when the coordinate system does not provide a regular description at the event horizon, an additive constant $T$ needs to be included in the HJ ansatz, $I = - \omega t + W(x^i) + T$, where $x^i$ denotes spatial coordinates and $T$ is used to fix the imaginary contribution associated with the time coordinate, thereby ensuring the correct normalization of the tunneling probabilities. Then, neglecting back-reaction effects, Hawking temperature $T_H$ is determined by $2{\rm Im} I \approx \omega/T_H$, where the imaginary part of the semiclassical action is obtained by evaluating the integral along a complex contour.
\section{The quantum fields in the Standard Model and Beyond} \label{sec: quantum-fields}
In this section, barred Greek indices (e.g., $\bar{\mu}$) refer to components defined in Minkowski spacetime.

Before deriving the Hawking spectrum, we briefly review the properties of fundamental particles observed in terrestrial laboratories, namely, those of the SM, as well as some well-motivated extensions, including supersymmetry and other scenarios beyond the SM.
\subsection{The Standard Model of particle physics} \label{sec: quantum-fields-SM}
The SM of particle physics is a gauge field theory described by the gauge group $SU(3)_c \times SU(2)_L \times U(1)_Y$, where $SU(3)_c$ is the color group associated with the strong interaction, and $SU(2)_L \times U(1)_Y$ is the electroweak part.
For the electroweak sector, the fields $B_{\bar\mu}$ and $W_{\bar\mu}^{\bar{a}}$ (with $\bar{a} = 1, 2, 3$) are the gauge fields associated with $U(1)_Y$ and $SU(2)_L$, respectively. In the unitary gauge for the Higgs field, all the unphysical degrees of freedom are removed from the spectrum. Then, after spontaneous symmetry breaking, $SU(2)_L \times U(1)_Y \rightarrow U(1)_{EM}$, and neglecting gauge interactions, the freely propagating gauge-boson sector can be described by the quadratic Lagrangian
    \begin{equation}
    \begin{aligned}
        {\cal L}_{AWZ} =& \frac{1}{2} A_{\bar\mu} D^{\bar{\mu} \bar{\nu}} A_{\bar\nu} + \frac{1}{2} Z_{\bar\mu} D^{\bar{\mu} \bar{\nu}} Z_{\bar\nu} + W_{\bar\mu}^{+} D^{\bar{\mu} \bar{\nu}} W_{\bar\nu}^{-}  \\ 
        &+ \frac{1}{2} m_Z^2 Z_{\bar\mu} Z^{\bar\mu} + m_W^2 W_{\bar\mu}^{+} W^{- \bar{\mu}}  \, , \label{Electroweak}
    \end{aligned}
    \end{equation}
where $
D^{\bar{\mu}\bar{\nu}} = \Box \eta^{\bar{\mu} \bar{\nu}} - (1 - \xi^{-1}) \partial^{\bar\mu} \partial^{\bar\nu}$ and the physical masses are $m_W = c_w m_Z$, with the short-notation: $c_w = \cos \theta_w, s_w = \sin \theta_w$. Here $\theta_w$ is the \emph{Weinberg's angle}. The physical electroweak gauge bosons are defined via
    \begin{align}
        W_{\bar\mu}^{+} &= \frac{1}{\sqrt{2}} (A_{\bar\mu}^1 - i A_{\bar\mu}^2), &
        W_{\bar\mu}^{-} &= \frac{1}{\sqrt{2}} (A_{\bar\mu}^1 + i A_{\bar\mu}^2) \, , \\
        Z_{\bar\mu} &= c_w A_{\bar\mu}^3 - s_w B_{\bar\mu} \, , &
        A_{\bar\mu} &= s_w A_{\bar\mu}^3 + c_w B_{\bar\mu} \, .
    \end{align}
Note that, in the Feynman gauge, $\xi=1$, the gauge boson sector can be written in a Proca-like form, which is particularly convenient for obtaining well-defined wave equations. In this gauge, each component of the free gauge fields satisfies a Klein–Gordon–type equation,
    \begin{equation}
        (\Box + M^2){\cal A}_{\bar\mu} = 0 \, , 
    \end{equation}
where the mass parameter is $M = 0, m_z, m_w$ for ${\cal A}_{\bar\mu} = A_{\bar\mu}, Z_{\bar\mu}, W_{\bar\mu}^{\pm}$, which describes the photon, $Z$-boson, and charged $W$-bosons, respectively. This form arises after gauge fixing, with the self-interactions among the gauge bosons neglected in order to describe free particles. It should therefore not be interpreted as reducing the full interacting gauge dynamics to that of scalar fields. Moreover, the physical degrees of freedom are selected by imposing the transversality condition
    \begin{equation}
        \partial^{\bar\mu} {\cal A}_{\bar\mu} = 0 \, ,
    \end{equation}
which eliminates unphysical polarizations.

On the other hand, the Quantum Chromodynamics (QCD) Lagrangian with $f$ quark flavors is written as
    \begin{equation}
        {\cal L}_{\rm QCD} = - \frac{1}{4} F_{\bar{\mu} \bar{\nu}}^a F^{a \bar{\mu}\bar{\nu}} + \sum_{f}\left(i \tilde{\gamma}^{\bar\mu} D_{s,\bar\mu} - m_f \right)\psi_f \, ,
    \end{equation}
where the color field strength is $F_{\bar{\mu} \bar{\nu}}^a = \partial_{\bar\mu} A_{\bar\nu}^a - \partial_{\bar\nu} A_{\bar\mu}^a + g_s f^{abc} A_{\bar\mu}^b A_{\bar\nu}^c$, with the non-abelian fields $A_{\bar\mu}^a$, where $g_s$ denotes the strong coupling and $f^{abc}$ the structure constant of $SU(3)_c$. The covariant gauge derivative is
    \begin{equation}
        D_{s,\bar\mu} = \partial_{\bar\mu} - ig_s T^a A_{\bar\mu}^a \, . 
    \end{equation}
where $T^a$ are the generators of 
$SU(3)_c$, with $a=1,2,\dots,8$.

Although gluons and quarks are not observed as free asymptotic states due to confinement below the energy scale $\Lambda_{\rm QCD}$, they are fundamental degrees of freedom of QCD and play a crucial role in the dynamics. Physical observables are described in terms of color-singlet states (hadrons), which arise from quarks and gluons through hadronization.

To describe the sector of quarks under a background with electric charge, we consider the Dirac equation for a particle of mass $m_f$ with a fermion field $\psi_f$;
    \begin{equation}
        \left(i \tilde{\gamma}^{\bar\mu} D_{\bar\mu} - m_f\right) \psi_f = 0 \, . 
        \label{D-fermions}
    \end{equation}
where $\tilde{\gamma}^{\bar\mu}$ are the corresponding Dirac matrices. The minimal coupling to the electromagnetic field is introduced via the covariant derivative
$D_{\bar\mu} = \partial_{\bar\mu}  - iq A_{\bar\mu} - ig_s T^a A_{\bar\mu}^a$, which describes the interaction of a charged fermion (with charge $q$) with the electromagnetic potential $A_{\bar\mu}$.

Without loss of generality, we can consider a similar Dirac equation~\eqref{D-fermions} for the SM leptons, such as electrons, positrons, muons, etc.

Note that the equation~\eqref{D-fermions} should actually run over all gauge group couplings; however, for the goal of our discussion throughout this paper, it will suffice to consider such a general form, since we will focus mainly on freely propagating particles on gravitational backgrounds.

\subsection{Supergravity} \label{subsec:Supergravity}
A supergravity theory is intrinsically nonlinear and, therefore, describes an interacting field theory. It necessarily includes the gravitational (gauge) multiplet and may also contain additional matter multiplets consistent with the underlying supersymmetry algebra. In the simplest case of ${\cal N}=1$ supergravity in four spacetime dimensions, the gauge multiplet comprises only the graviton and a single Majorana gravitino~\cite{Freedman2012supergravity}.

Instead of treating the exact supergravity theory, we only focus on a linearized version, where the graviton is described by a small perturbation $|\kappa h_{\bar{\mu}\bar{\nu}}| \ll 1$ around the Minkowski spacetime
    \begin{equation}
        g_{\bar{\mu}\bar{\nu}} = \eta_{\bar{\mu}\bar{\nu}} + \kappa h_{\bar{\mu}\bar{\nu}} \quad \text{with} \quad \kappa = \sqrt{8\pi G_N} \, ,
    \end{equation}
where $G_N$ is the gravitational constant.

To obtain the description at the Lagrangian level, one can expand the Einstein-Hilbert Lagrangian to second order in the perturbation, taking into account the Minkowski background. It is worth noticing that this prescription is equivalent to the Fierz-Pauli action up to total derivatives in the neutral massless case~\cite{Fierz1939relativistic}.

To describe a massless graviton, it is useful to use an auxiliary field $\bar{h}_{\mu\nu} = h_{\bar{\mu}\bar{\nu}} - \frac{1}{2}\eta_{\bar{\mu}{\nu}} h$, where $h$ is the trace of $h_{\bar{\mu}\bar{\nu}}$, after using the de Donder gauge $\partial^\mu \bar{h}_{\bar{\mu}\bar{\nu}} = 0$, we obtain a wave equation $\Box \bar{h}_{\bar{\mu}\bar{\nu}} =0$.

Since the linearized Einstein equations are invariant under the gauge transformations $\delta \bar{h}_{\bar{\mu}\bar{\nu}} = \partial_{\bar\mu} \xi_{\bar\nu} + \partial_{\bar\nu} \xi_{\bar\mu} - \eta_{\bar{\mu}\bar{\nu}} \partial_{\bar\lambda} \xi^{\bar\lambda}$ with $\xi_{\bar\mu}$ an arbitrary vector field, it follows that $\delta (\partial_{\bar\mu} \bar{h}_{\bar{\mu}\bar{\nu}}) = \Box \xi_{\bar\nu}$~\cite{Van1981supergravity}. This shows that the gauge condition does not completely fix all degrees of freedom, leaving a residual gauge freedom. This issue can be ruled out in the case of a massive graviton in a curved background, which we discuss later.

On the other hand, the gravitino is described by the Rarita–Schwinger theory~\cite{Rarita1941theory}; its equation of motion can be reduced to an effective Dirac equation by constructing appropriate projectors via a Gram–Schmidt procedure. These projectors isolate the physical spin-$3/2$ degrees of freedom and remove non-physical components (see, e.g., ~\cite{Freedman2012supergravity} or Appendix A of Ref.~\cite{P2025cpt} for a detailed derivation),
    \begin{equation}
        \left( i\tilde{\gamma}^{\bar\mu} \partial_{\bar\mu} - m_g\right) \boldsymbol{\psi}_{\bar\lambda} = 0 \, .
    \end{equation}
that satisfies $\tilde{\gamma}^{\bar\mu}\boldsymbol{\psi}_{\bar\mu} =0$. 

If one considers a gravitino minimally coupled to electromagnetism, analogously to the \ipc{SM} fermions in Eq.~\eqref{D-fermions}, i.e., carrying electric charge, the resulting equations of motion exhibit superluminal propagation, this is the so-called Velo-Zwanziger problem~\cite{Velo1969propagation} and non-causal propagation~\cite{Johnson1961inconsistency}. It has therefore been suggested that introducing non-minimal couplings to the electromagnetic field can resolve this pathology in a constant external field strength $F_{\mu\nu}$~\cite{Ferrara1992g, Porrati2009causal}.
\subsection{Higher-spin fields} \label{subsec: higher-spin}
In many extensions to higher-spin fields, a Proca Lagrangian is often used to describe a massive spin-$1$ vector field. However, the presence of the mass term explicitly breaks the underlying $U(1)$ gauge invariance. The Lagrangian is given by
    \begin{equation}
        {\cal L}_V = - \frac{1}{4} F_{\bar{\mu}\bar{\nu}}^V {F^{V}}^{\bar{\mu}\bar{\nu}} + \frac{1}{2} m_V^2 V_{\bar\mu} V^{\bar\mu} \, , \label{Proca Lagrangian}
    \end{equation}
where $F_{\bar{\mu}\bar{\nu}}^{V} = \partial_{\bar\mu} V_{\bar\nu} - \partial_{\bar\nu} V_{\bar\mu}$.

Here, the non-null mass term naturally gives us the Lorentz condition $\partial^{\bar\mu} V_{\bar\mu} =0$.

This description should be understood either as an effective theory or as emerging from a gauge-invariant framework after spontaneous symmetry breaking, as realized in the SM via the Higgs mechanism. Alternatively, gauge invariance may be restored through the Stückelberg mechanism by introducing compensating scalar fields; however, its non-abelian generalization is highly nontrivial and typically encounters difficulties associated with unitarity and renormalizability~\cite{Ruegg2004stueckelberg}.

To generalize relativistic wave equations to higher-spin fields with $s>2$, one may consider an extension of the Klein–Gordon-like equation acting on massive higher-spin fields. These fields are described by totally symmetric, (gamma-)traceless rank-$s$ tensors $\phi_{\bar{\mu}_1 \cdots \bar{\mu}_s}$. 

Lorentz invariance implies that $\phi_{\bar{\mu}_1 \cdots \bar{\mu}_s}$ contains more components than the physical number of degrees of freedom, namely $2s+1$. The redundant components must be removed by imposing the transversality constraint
    \begin{equation}
        \partial^{\bar{\nu}} \phi_{\bar{\nu} \bar{\mu}_2 \cdots \bar{\mu}_s} = 0 \, .
    \end{equation}
A consistent implementation of this constraint requires the introduction of a set of auxiliary fields~\cite{Singh1974lagrangian, Singh1974lagrangian2}.

However, this framework is essentially restricted to free fields. The inclusion of interactions, particularly with gauge fields or gravity, generically leads to well-known consistency issues. In particular, the $\phi_{\bar{\mu}_1 \cdots \bar{\mu}s}$ fields typically require gauge symmetries to eliminate unphysical degrees of freedom. At the same time, naive minimal couplings can violate causality, unitarity, or even Lorentz invariance. In some cases, these problems can be mitigated by introducing appropriate non-minimal interaction terms, which restore causality and unitarity of the theory~\cite{Porrati2009causal, Cuccieri1995tree}.

\section{Quantum field theory in curved space time} \label{sec: quantum-fields-curved}
In this section, we present the elements for working in curved spacetime; here, the Greek indices refer to curved spacetime, and we use the West Coast convention in the metrics, but the results will be independent of this choice.

We consider a BH spacetime in ${\rm D}=4$ dimensions, described by a metric $g_{\mu\nu}$ that solves the Einstein–Maxwell equations in vacuum. According to the BH uniqueness (no-hair) theorems, stationary asymptotically flat solutions of the Einstein–Maxwell equations are fully characterized by their mass, angular momentum, and electric charge, and are described by the KN family (see~\cite{Chrusciel2012stationary} and references therein).

In the Boyer-Lindquist (BL) coordinates, the KN black hole is described by~\cite{Kerner2006tunnelling}
    \begin{equation}
    \begin{aligned}
        ds^2 =& f(r,\theta) dt^2 + 2H(r,\theta) dt d\varphi - \frac{1}{G(r,\theta)} dr^2 \\
        &- \Sigma(r,\theta) d\theta^2 - K(r, \theta) d\varphi^2  \, .
    \end{aligned} \label{KN-metric}
    \end{equation}
Here, the metric elements are:
    \begin{equation}
    \begin{aligned}
        f(r, \theta) &= \frac{\Delta(r)-a^2 \sin^2 \theta}{r^2 + a^2 \cos^2\theta} \, ,\\
        H(r, \theta) &= \frac{a\sin^2 \theta[(r^2 + a^2) - \Delta(r)]}{r^2 + a^2 \cos^2\theta} \, ,\\
        G(r, \theta) &= \frac{\Delta(r)}{r^2 + a^2 \cos^2\theta} \, ,\\
        \Sigma(r, \theta) &= r^2 + a^2 \cos^2\theta \,,\\
        K(r, \theta) &= \frac{\sin^2 \theta[(r^2 + a^2)^2 - a^2 \Delta(r) \sin^2 \theta]}{r^2 + a^2 \cos^2\theta} \, ,
    \end{aligned} \label{KN-LE}
    \end{equation}
where $\Delta(r)=(r-r_{-})(r-r_{+})$ is determined by the outer ($+$) and inner ($-$) horizons,
    \begin{equation}
        r_{\pm}= M \pm \sqrt{M^{2}-a^{2}-Q^{2}} \, . \label{Horizonts}
    \end{equation}
Here, $M$ is the black hole mass, $a$ determines the angular moment $J=Ma$, and $Q$ is the electrical charge that determines the $4$-potential,
    \begin{equation}
        A_\mu dx^\mu = -\frac{Qr}{\Sigma(r, \theta)} (dt - a\sin^2 d\theta) \, . \label{A-potential}
    \end{equation}
The following wave equations will be formulated on this background spacetime (where $R=0$), and the corresponding electromagnetic interaction terms for SM and beyond the SM particles will be introduced below.
\subsection{Bosons in a curved spacetime}
In the SM, there exists a scalar field corresponding to the neutral, massive Higgs boson. At leading order, with its interactions with other SM particles neglected, it can be treated as a free massive scalar field propagating on the background geometry. In a KN spacetime, it therefore satisfies the Klein–Gordon (KG) equation in curved spacetime,
    \begin{equation}
        \frac{1}{\sqrt{-g}} \partial_\mu (\sqrt{-g}g^{\mu\nu} \partial_\nu H) + \frac{m_H^2}{\hslash^2} H = 0 \, .
        \label{KG-Higgs}
    \end{equation}
This corresponds to the neutral limit of a complex scalar field (see Eq.~\eqref{S-scalar}), obtained by taking $q \to 0$, and $H$ now represents a real scalar field.

For the $Z$-boson, we consider the action in a KN spacetime; 
    \begin{equation}
        {\cal S}_{Z}= \int dx^4 \sqrt{-g}  \left( - \frac{1}{4} Z^{\mu\nu} Z_{\mu\nu} + \frac{1}{2}\frac{m_Z^2}{\hslash^2} Z^\mu Z_\mu \right) \, ,
    \end{equation}
where $Z_{\mu\nu} = \nabla_\mu Z_\nu - \nabla_\nu Z_\mu$, with $\nabla_\mu Z_\nu = \partial_\mu Z_\nu - \Gamma_{\mu\nu}^\lambda Z_\lambda$. 

As these $Z$-bosons are neutral, the motion equation that applies to both Kerr and KN spacetime is
    \begin{equation}
        \nabla_\mu Z^{\mu\nu} + \frac{m_Z^2}{\hslash^2} Z^\nu = 0 \, . \label{KG-Z}
    \end{equation}
Similarly, for gluons, which are massless particles without electric charge above the QCD confinement scale $\Lambda_{\rm QCD} \approx 0.17 \, {\rm GeV}$: 
    \begin{equation}
        \nabla_\mu F^{\mu\nu a} = 0 \, , \label{KG-gluon}
    \end{equation}
Here $F_{\mu\nu}^{a} = \nabla_\mu F_\nu^a - \nabla_\nu F_\mu^a$, with $\nabla_\mu F_\nu^a = \partial_\mu F_\nu^a - \Gamma_{\mu\nu}^\lambda F_\lambda^a$. 

On the other hand, for photons, we consider perturbations that describe the field of photons and another for the background; in this case, we can write
    \begin{equation}
        {\cal F}_{\mu\nu} = F_{\mu\nu} + f_{\mu\nu} \, ,
    \end{equation}
where $F_{\mu\nu} = \nabla_\mu A_\nu - \nabla_\nu A_\mu$, and the electromagnetic tensor for the photon is $f_{\mu\nu} = \nabla_\mu {\rm a}_\nu - \nabla_\nu {\rm a}_\mu$. In this scenario, the motion equation for the photons under the KN background is 
    \begin{equation}
        \nabla_\mu f^{\mu\nu} =0 \, . \label{KN-photons}
    \end{equation}
For charged $W$ bosons, an additional non-minimal coupling term is required to account for the gyromagnetic moment of charged SM particles. The resulting equation of motion is given by (see Appendix~\ref{AppC}):
    \begin{equation}
        {\cal D}_\mu^{\pm} F^{\pm \mu\nu} + \frac{m_W^2}{\hslash^2} W^{\pm \nu} \pm \frac{i}{\hslash} e F^{\nu\mu} W_\mu^{\pm} =0 \, , \label{KG-W}
    \end{equation}
where $F_{\mu\nu}^{\pm} = {\cal D}_\mu^{\pm} W^{\pm}_\nu - {\cal D}_\nu^{\pm} W^{\pm}_\mu$, with $\nabla_\mu W^{\pm}_\nu = \partial_\mu^{\pm} W^{\pm}_\nu - \Gamma_{\mu\nu}^\lambda W^{\pm}_\lambda$, and ${\cal D}_\mu^{\pm} = \nabla_\mu \pm ie A_\mu/\hslash$.

A particularly instructive case is the hypothetical graviton. As \ipc{we} discussed in Subsec.~\ref{subsec:Supergravity}, a massless graviton can be consistently described as a perturbation around Minkowski spacetime. However, in a curved background, the identification of the physical degrees of freedom is more subtle. In such cases, it is often more convenient to consider a massive spin-$2$ field. Following the prescription of Ref.~\cite{Buchbinder2000equations}, the equation of motion for a massive graviton described by the field $h_{\mu\nu}$ is
    \begin{equation}
        \nabla^\rho \nabla_\rho h_{\mu\nu} + 2 {{{R^\alpha}_\mu}^\beta}_\nu h_{\alpha\beta} + \frac{m^2}{\hslash^2} h_{\mu\nu} =0 \, , \label{KN-graviton}
    \end{equation}
supplemented by the constraints
    \begin{equation}
        h = 0 \quad \text{and} \quad \nabla^\mu h_{\mu\nu} = 0 \, , \label{graviton-const}
    \end{equation}
which enforce the correct number of propagating degrees of freedom. In addition, further relations arise from the time component of the divergence constraint and from the absence of independent time derivatives for certain components (see Ref.~\cite{Buchbinder2000equations} for details). 

Taking the massless limit in this framework—and more generally in curved backgrounds—leads to the well-known van Dam–Veltman–Zakharov (vDVZ) discontinuity~\cite{Van1970massive, Zakharov1969linearized, Porrati2002fully}. Consequently, we will instead consider a massive graviton with a small but nonzero mass $m$. 
\subsection{Fermions in a curved spacetime}
To describe fermions in curved spacetime, we cannot define them directly in terms of general coordinate transformations, since there is no finite-dimensional spinor representation of $GL(4)$.
Instead, we introduce a tetrad (vierbein) field $e_\mu^{\bar\nu}$, which defines a local Lorentz frame at each spacetime point. The tetrad relates the curved spacetime metric $g_{\mu\nu}$ to the Minkowski metric $\eta_{\bar{\mu}\bar{\nu}}$ through
    \begin{equation}
        g_{\mu\nu} = e_\mu^{\bar\mu} e_\nu^{\bar\nu} \eta_{\bar{\mu}\bar{\nu}} \, .
    \end{equation}
In this local inertial frame, spinors transform under the usual finite-dimensional representations of the Lorentz group $SO(1,3)$, making it possible to formulate fermionic theories in curved backgrounds consistently. The ordinary derivative must then be replaced by a covariant derivative containing the spin connection $\omega_\mu^{\bar{\mu}\bar{\nu}}$,
    \begin{equation}
        {\cal D}_\mu = \partial_\mu - i\frac{q}{\hslash}A_\mu + \Omega_\mu \quad \text{with} \quad \Omega_\mu := \frac{1}{4} \omega_\mu^{\bar{\mu}\bar{\nu}} \tilde{\gamma}_{\bar{\mu}\bar{\nu}} \, ,
    \end{equation}
where $\tilde{\gamma}_{\bar{\mu}\bar{\nu}}=[\tilde{\gamma}_{\bar{\mu}}, \tilde{\gamma}_{\bar{\nu}}]/2$ is defined in terms of the gamma matrices $\tilde{\gamma}^{\bar{\mu}}$, this also determines $\gamma^\mu = e^\mu_{\bar\nu} \tilde{\gamma}^{\bar\nu}$. We employ the minimal coupling to the vector field $A_\mu$ to describe fermions with electric charge $q$.

Particularly for spin-$1/2$ fermions,
    \begin{equation}
        \left(i \gamma^\mu {\cal D}_\mu - \frac{m_f}{\hslash} \right) \psi = 0 \, . \label{Motion.Eq.D}
    \end{equation}
In a minimal supergravity scenario, one must include a gravitino; however, if it is assumed to be minimally coupled to electromagnetism, the resulting description is not consistent. Instead, we assume an electrically neutral gravitino that is described by extended Rarita-Schwinger theory~\cite{Amsel2009supergravity} using the field $\Psi_\lambda$, then the generalization is given by 
    \begin{equation}
        \left(i \gamma^\mu {\cal D}_\mu - \frac{m_g}{\hslash} \right) \Psi_\lambda = 0 \, ,
        \label{Motion.Eq.RS}
    \end{equation}
with the condition $\gamma^\mu \Psi_\mu = 0$.

In flat spacetime, a KG-like equation for fermions can be obtained by acting the operator $\left(i \tilde{\gamma}^{\bar\mu} D_{\bar\mu} + m_f\right)$ on the Dirac equation in Eq.~\eqref{D-fermions}. Analogously, in curved spacetime, the same procedure can be applied to derive a KG-like equation for fermionic fields. In particular, for a spin-$1/2$ fermion with electric charge $q$ and a neutral ($q=0$) spin-$3/2$ fermion, the corresponding equations can be written as ($\hslash \neq 1$), respectively:
\begin{widetext}
    \begin{align}
        (i \gamma^{\mu} {\cal D}_{\mu} + m_f)(i \gamma^{\nu} {\cal D}_{\nu} - m_f) \psi &= \left[- g^{\mu\nu} {\cal D}_{\mu} {\cal D}_{\nu} - \frac{1}{8} \gamma^{\mu\nu} R_{\mu\nu\rho\sigma} \gamma^{\rho\sigma} - \frac{iq}{2 \hslash}\gamma^{\mu\nu}F_{\mu\nu} - \frac{m_f^2}{\hslash^2} \right]\psi =0 \, , \label{KG-fermions12} \\
        (i \gamma^{\mu} {\cal D}_{\mu} + m_f)(i \gamma^{\nu} {\cal D}_{\nu} - m_f) \Psi_\lambda &= \left[ - g^{\mu\nu}{\cal D}_{\mu} {\cal D}_{\nu} - \frac{1}{8} \gamma^{\mu\nu} R_{\mu\nu\rho\sigma} \gamma^{\rho\sigma} - \frac{m_g^2}{\hslash^2} \right]\Psi_\lambda - \frac{1}{2}\gamma^{\mu\nu} {{R_\lambda}^\sigma}_{\mu\nu} \Psi_\sigma \, , \label{KG-fermions32}
    \end{align}
\end{widetext}
Here, $\gamma^{\mu\nu} = [\gamma^\mu, \gamma^\nu]/2$.
\section{The separability of Hamilton-Jacobi equation and Hawking radiation} \label{sec:HJ and HR}
In the semiclassical limit, wave equations reproduce the classical equations of motion. Hence, by this correspondence principle, any separability property of the KG equation implies separability of the associated HJ equation~\footnote{This observation has been used as a guiding principle in solving Einstein’s equations, leading to the derivation of the Kerr and KN spacetimes~\cite{Carter1968global, Carter2009republication}.}. In this part, we exploit this statement to derive the Hawking spectrum for BHs that satisfy the no-hair theorem~\cite{Israel1967event, Israel1968event, carter1971axisymmetric, Robinson1975uniqueness, Mazur1982proof}: the most general case is a KN black hole.
\subsection{The Klein-Gordon equation case}
To elucidate the general characteristics of our proposal, we consider a complex scalar field $\Phi$ in a KN background, described by the KG equation as
    \begin{equation}
        g^{\mu\nu} D_\mu D_\nu \Phi + \frac{m_\phi^2}{\hslash^2} \Phi=0 \, . \label{KG}
    \end{equation}
where $D_\mu = \nabla_\mu - iq A_\mu/ \hslash$ is the covariant gauge derivative.
    
To solve the equation~\eqref{KG}, we use the ansatz 
    \begin{equation}
    \Phi(t,\theta,\varphi,r) = \phi \exp\left( \frac{i}{\hslash} I(t,\theta,\varphi,r) \right)  \, ,
    \end{equation}
with $\phi$ a complex constant.

After doing the classical approximation $\hslash \to 0$, we obtain the well-known form of the HJ equation
    \begin{equation}
        g^{\mu\nu}(\partial_\mu I - qA_\mu)(\partial_\nu I - qA_\nu ) - m_\Phi^2 =0 \, . \label{HJ-scalar}
    \end{equation}
The Eq.~\eqref{HJ-scalar} in BL coordinates can be solved by separation of variables. To this end, we propose the ansatz
    \begin{equation}
        I = -\omega t + W(r,\theta) + \ell \varphi + T \, . \label{s-action-ansatz}
    \end{equation}
Here, $\omega = - \partial_t I$ is the conserved quantity associated with the stationarity of the spacetime, $\ell$ is the conserved azimuthal angular momentum associated with axial symmetry (invariance under rotations in $\varphi$). Here, we have added a constant $T$ because the BL coordinates do not provide a regular description across the event horizon.

Notice that we have kept $W(r,\theta)$ without further separation into radial and angular parts. This choice is motivated by the fact that, in the null-geodesic method, the metric functions~\eqref{KN-LE} depend explicitly on the polar angle $\theta$. As argued in the literature, \emph{the presence of $\theta$-dependence implies that we can no longer just look at a generic spherical wave. Instead, one considers the emission of particles along trajectories at a fixed angle $\theta = \theta_0$, effectively analyzing rings of emitted particles rather than spherical shells}~\cite{Kerner2006tunnelling}.

The previous observation has been translated into the standard complex-path approach, and, to date, the solution of the radial part of this equation relies on an approximation near the event horizon~\cite{Kerner2006tunnelling}, fixing $\theta=\theta_0$ as $r \to r_{+}$. However, by construction, the KN spacetime admits a separable HJ equation in the BL coordinates~\cite{Carter1968global}, so an angular approximation is not necessary. 

Following the detailed steps presented in Appendix~\ref{AppA}, and using the ansatz~\eqref{s-action-ansatz}, with $W(r,\theta) = {\cal R}(r)+ {\cal T}(\theta)$, we can write the HJ equation~\eqref{HJ-scalar} as:
\begin{widetext}
\begin{equation}
    \frac{[(r^2 + a^2) \omega - a\ell - q Qr]^2}{\Delta(r)} - \left[ a\omega\sin\theta - \frac{\ell}{\sin\theta} \right]^2 - \left(\frac{d{\cal T}}{d\theta}\right)^2 - \Delta(r)\left(\frac{d{\cal R}}{dr} \right)^2 - (r^2 +a^2 \cos^2\theta) m_\Phi^2 = 0  \, . \label{HJ-separable}
\end{equation}
\end{widetext}
We introduce $\lambda$ as the separation constant in the HJ equation~\eqref{HJ-separable}, separating the $(t,r)$ and $(\theta,\varphi)$ coordinates (see details in Appendix~\ref{AppA}). This shows that the imaginary part of the semiclassical action can be determined without fixing any polar angle.

To better understand the separability of the HJ equation~\eqref{HJ-separable}, we pointed out Noether’s theorem, which states that every continuous symmetry of the action is associated with a conserved quantity, provided the equations of motion are satisfied. This principle extends beyond classical physics: in quantum field theory, analogous statements are encoded in the Ward identities, which relate symmetries to quantum-mechanical conservation laws.

In general relativity, continuous spacetime symmetries are generated by Killing vector fields, and they lead to conserved quantities along geodesics, such as energy and angular momentum. Nevertheless, not all symmetries are manifest. There exist additional, less obvious symmetries known as \emph{hidden symmetries}. These do not arise from Killing vectors but instead from higher-rank geometric objects, such as \emph{Killing tensors}~\footnote{These tensors are sometimes referred to as Stäckel--Killing tensors, since Stäckel, a contemporary of Killing, was in fact the first to study the type of Hamilton--Jacobi separability that leads to the existence of such tensors~\cite{Carter2009republication}.}.

Hidden symmetries give rise to conserved quantities along particle trajectories, ensuring the integrability of the equations of motion. Specifically, second-rank Killing tensors generate conserved quantities that are quadratic in the particle momenta. Before presenting our result, it is worth recalling that, after the discovery by Carter of an unexpected constant of motion in the Kerr spacetime~\cite{Carter1968global}, Walker and Penrose showed that this conserved quantity originates from a rank-two Killing tensor~\cite{Walker1970quadratic}. In the present case, the Killing tensor $k_{\mu\nu}$ provides the constant of motion
    \begin{equation}
        {\cal K}= k_{\mu\nu} p^\mu p^\nu \, . \label{Carter-cte}
    \end{equation}
where $p^\mu$ is the tangent vector (four-momentum) to the particle’s geodesic. 

For the KN spacetime, the existence of such a hidden symmetry leads to the Carter constant ${\cal K}$. In the asymptotically flat region, this conserved quantity admits an interpretation as a generalized notion of total angular momentum for a particle or field, extending the standard conserved quantities associated with manifest rotational symmetry~\cite{De1999meaning}.

One can show that $\lambda = {\cal K}$ (see Sec.~8 of Ref.~\cite {Carter2009republication}). Therefore, the solutions of the radial part can be determined as follows:
    \begin{equation}
        \frac{d{\cal R}}{dr} = \pm \frac{\sqrt{(r^2 +a^2)^2\left[ \omega -\Omega \ell - q\Phi_r \right]^2 - \Delta(r) F(r)}}{\Delta(r)} \, ,
    \end{equation}
where $\Omega = a/(r^2 +a^2)$, $\Phi_r= Qr/(r^2 + a^2)$, and $F(r)=r^2  m_\Phi^2 + {\cal K}$. 

Now, we need to determine the radial contribution, so we integrate around the event horizon $r_H = r_{+}$. First, consider the expansion:
    \begin{equation}
        \Delta(r) = \Delta(r_H) + \Delta^\prime (r_H) (r-r_H) + \cdots \, .
    \end{equation}
Here $\Delta^\prime (r) = 2(r-M)$, 
    
Using $r-r_H = \epsilon e^{i\vartheta}$, $\vartheta \in (\pi, 2\pi)$: $dr = i\epsilon e^{i\vartheta} d\vartheta$
    \begin{equation}
        \int_{r_H - \epsilon}^{r_H + \epsilon} \frac{dr}{(r-r_H)} = \int_\pi^{2 \pi} i d\vartheta = i \pi \, .
    \end{equation}
Therefore, we choose the sing $+$ for outgoing particles and $-$ for incoming particles, we obtain that the contributions:
    \begin{equation}
        {\cal R}_{\pm}= \pm i\pi \frac{(r_H^2 + a^2) \omega_{\rm eff}}{2(r_H - M)} \, .
    \end{equation}
where the effective energy of the particle has a contribution from the angular velocity of the horizon $\Omega_H$ and another from the electric potential at the horizon $\Phi_H$, and is $\omega_{\rm eff} = \omega -\Omega_H \ell - q\Phi_H$.

As we explained in Sec.~\ref{sec:Complex-path-HR}, and write explicitly in Eq.~\eqref{s-action-ansatz}, we need to add an extra constant $T$ because the BL coordinates do not smoothly describe infalling trajectories across the event horizon; here $T=-{\cal R}_{-}$ to describe $P[in] =1$. Then, the tunneling rate is
    \begin{equation}
        \Gamma = \exp \left\{- \frac{ 2(r_H^2 + a^2) \pi \omega_{\rm eff}}{(r_H - M)}\right\} = \exp\left\{-\frac{\omega_{\rm eff}}{T_H}\right\} \, .
    \end{equation}
Therefore, we obtain the correct Hawking temperature
    \begin{equation}
        T_H = \frac{(r_H-M)}{2 \pi (r_H^2 + a^2)} = \frac{(r_{+}- r_{-})}{4\pi (r_{+}^2 + a^2)} \, . \label{T_H}
    \end{equation}
Here, we do not consider energy, angular momentum, and charge conservation, which generalizes Eq.~\eqref{radii} that implies energy conservation in the BH-particle system, sometimes called \emph{self-gravitation}~\cite{Kraus1995self, Parikh2004energy}. In the absence of back-reaction, particles created just inside the horizon would only need to tunnel across an infinitesimal separation, implying the absence of a potential barrier. However, once back-reaction is taken into account, the horizon radius shifts, and the resulting finite separation between the initial and final radii defines a classically forbidden region, i.e., the tunneling barrier.

When back-reaction is taken into account, we obtain $2Im I = - \Delta S_{B-H}$, where $\Delta S_{B-H}$ denotes the change in the Bekenstein--Hawking entropy (see App.~\ref{AppB} for a derivation):
    \begin{equation}
        \Delta S_{B-H} = [S(M-\omega, J-\ell , Q-q) - S(M, J, Q)] \, ,
    \end{equation}
where the usual Bekenstein--Hawking entropy is given by
    \begin{equation}
        S(M, J, Q) = \pi \left[ M + \sqrt{M^2 - \frac{J^2}{M^2}-Q^2} + \frac{J^2}{M^2}\right] \, . \label{BH-entropy}
    \end{equation}
\section{Hawking spectrum for the standard model particles} \label{sec: HR for SM}
The idea that HR is universal for all particles and spins has been explored~\cite{Erbin2018universality}; however, there is no explicit explanation for why the mechanism works well for rotating BHs. In the following, we present an explicit derivation of HR for SM particles in the KN background. In the following WKB ansatz, we omit the explicit dependence on $t$, $\theta$, $\varphi$, and $r$ in most cases ($I= I(t,\theta,\varphi,r)$); however, it is understood to be present in the subsequent computations.
\subsection{Higgs boson case}
We consider the KG-like equation~\eqref{KG-Higgs} for the Higgs boson. To solve the equation, we propose the WKB ansatz:
    \begin{equation}
        H = h \exp\left( \frac{i}{\hslash} I \right) \, ,
    \end{equation}
with $h$ a constant. 

Notice then that, as in the complex scalar case (see Sec.~\ref{sec:Complex-path-HR}), taking the  limit $\hslash \to 0$, we can write
    \begin{equation}
        \Upsilon(r,\theta)-m_H^2 =0 \quad \text{with} \quad \Upsilon(r,\theta) := g^{\mu\nu} \partial I \partial_\nu I \, . \label{Upsilon}
    \end{equation}
After multiplying \eqref{Upsilon} by $\Sigma(r,\theta)$, it is easy to show that we obtain the expression~\eqref{HJ-separable}, and therefore, we obtain the Hawking spectrum~\eqref{T_H} with $q=0$. The emission of such quanta is then controlled by the BH temperature, and is significant only when $T_H \gtrsim m_H$.
\subsection{Gauge particles tunneling}
We consider the motion equation for the $Z$-boson~\eqref{KG-Z}, using the WKB method, and we propose the ansatz 
    \begin{equation}
        Z_\lambda = z_\lambda \exp\left(\frac{i}{\hslash} I \right) \quad \text{with} \quad z_\lambda = z_\lambda (t,\theta, \varphi, r)\, .
    \end{equation}
After some algebra, taking the limit $\hslash \to 0$, we obtain
    \begin{equation}
        g^{\nu\lambda}[- \Upsilon(r,\theta) z_{\lambda} + ( g^{\mu\rho})  (\partial_\mu I)( z_{\rho} \partial_\lambda I) + m_Z^2  z_{\lambda} ] =  0 \, .
    \end{equation}
To obtain a well-defined wave equation, we consider the constraint
    \begin{equation}
        \nabla^\rho Z_\rho =0 \, , 
    \end{equation}
which at the limit $\hslash \to 0$ becomes to 
    \begin{equation}
        g^{\mu \rho} z_{\rho} \partial_\mu I =0 \, , \label{Z-constrain}
    \end{equation}
we obtain the expression
    \begin{equation}
        g^{\nu\lambda} [\Upsilon(r,\theta) - m_Z^2] z_{\lambda} = 0 \, .
    \end{equation}
As the non-trivial solution is obtained by taking the determinant of this system of equations equal to zero, $M(r,\theta)$, in this case, we obtain 
    \begin{equation}
        \det [M(r,\theta)] = [\Upsilon(r,\theta)-m_Z^2]^4  g^{rr} g^{\theta\theta} [g^{tt} g^{\varphi\varphi} - (g^{t\varphi})^2] =0 \, .
    \end{equation}
Therefore, the solution is $\Upsilon(r,\theta)-m_Z^2 = 0$, from which it is straightforward to obtain the Hawking temperature because it is the same condition as in the Higgs case.

Above the $\Lambda_{\rm QCD}$ confinement scale, the black hole can emit gluons, using the motion equation~\eqref{KG-gluon}, as for the $Z$-boson case, but now we propose the WKB solution,
    \begin{equation}
        A_\lambda^a = {\rm a}_\lambda^a \exp\left( \frac{i}{\hslash} I \right) \quad \text{with} \quad {\rm a}_\lambda^a= {\rm a}_\lambda^a(t,\theta, \varphi, r)\, .
    \end{equation}
In the classical limit $\hslash \to 0$, after some algebra, one obtains
    \begin{equation}
        g^{\mu\rho} g^{\nu\lambda} (\partial_\mu I)({\rm a}_{\lambda}^a \partial_\rho I - {\rm a}_{\rho}^a \partial_{\lambda} I) = 0 \, .
    \end{equation}
Again, to obtain a well-defined wave equation and apply the correspondence principle consistently, we use the gauge condition
    \begin{equation}
        \nabla^\rho A_\rho^a = 0 \Rightarrow g^{\mu\rho}{\rm a}_{\rho}^a \partial_\mu I = 0 \, .
    \end{equation}
With this condition, we reduce the motion equation to 
    \begin{equation}
        g^{\nu \lambda} \Upsilon(r,\theta) {\rm a}_{\lambda}^a =0 \, .
    \end{equation}
Here, the non-trivial solution satisfies $\Upsilon(r,\theta) =0$. As the result is mass independent, the Hawking spectrum coincides with that obtained in the previous cases.

Another simple case is to consider a photon; in such a case, the photon field must be considered as a perturbation of the electromagnetic field of the KN background and follows a motion equation~\eqref{KN-photons}; then we propose a WKB ansatz,
    \begin{equation}
        {\rm a}_\lambda = a_\lambda \exp\left( \frac{i}{\hslash} I\right) \quad \text{with} \quad a_\lambda = a_\lambda (t,\theta, \varphi, r) \,.
        \label{Photon-WKB}
    \end{equation}
Then, to zeroth order in $\hslash$, we obtain 
    \begin{equation}
        g^{\mu\rho} g^{\nu\lambda} (\partial_\mu I) ( a_{\lambda} \partial_\rho I - a_{\rho} \partial_\lambda I) =0 \, .
    \end{equation}
To consider a well-defined wave equation, one must use a gauge; in this case, we use the Lorentz gauge $\nabla^\rho A_\rho=0$, which gives us the constraint $g^{\mu\rho} a_{\rho} (\partial_\mu I) = 0$. Then the equation that one needs to solve is 
    \begin{equation}
        g^{\nu \lambda} \Upsilon(r,\theta) {\rm a}_{\lambda} =0 \, .
    \end{equation}
Notice that we recover the correct Hawking spectrum for the Abelian field. In the non-Abelian case (e.g., gluons), the genuinely non-Abelian contributions must be absent due to the lack of color charge; consequently, the horizon is sensitive only to the Abelian component.
So far, all cases have been straightforward to derive. We now turn to a more interesting scenario: the charged $W$ boson, which--owing to its electric charge--also exhibits a non-minimal coupling to the vector potential $A_\mu$. 

First, we consider the $W$ boson with charge $+e$, then in this case, 
    \begin{equation}
        W_\lambda^{+} = w_\lambda^{+} \exp\left( \frac{i}{\hslash} I \right) \quad \text{with} \quad  w_\lambda^{+}=  w_\lambda^{+}(t,\theta, \varphi, r)\, . \label{WKB-W}
    \end{equation}
In the classical limit, $\hslash \to 0$,
from Eq.~\eqref{KG-W}, we obtain (see details in Appendix~\ref{AppC});
    \begin{equation}
         g^{\nu\lambda}[ \Upsilon_{-e}(r,\theta) - m_W^2] w_{\lambda}^{+} + e g^{\mu\rho} g^{\nu\lambda}( \partial_\lambda I + eA_\lambda)w_{\rho}^{+} = 0 \, , \label{W-non-gauge}
    \end{equation}
where we have used (for $q=-e$):
    \begin{equation}
        \Upsilon_q(r,\theta) = g^{\mu\rho} (\partial_\mu I - qA_\mu) \left( \partial_\rho I - qA_\rho \right) \, . \label{Upsilon-q}
    \end{equation}
Due to the presence of electromagnetic interactions, it is necessary to impose an additional constraint in order to obtain a well-defined wave equation that propagates only the physical degrees of freedom, as follows
    \begin{equation}
        {\cal D}_{\mu}^{\pm} W^{\pm \mu} =0 \, .
    \end{equation}
This implies the following constraint
    \begin{equation}
        g^{\mu\rho} \left[ (\partial_\mu I) \pm e A_\mu  \right] w_{\rho}^{\pm} = 0 \, .
    \end{equation}
And then, the non-trivial solution that also applies for $W$ bosons with charge $-e$ is given by
    \begin{equation}
        \Upsilon_{\mp e}(r,\theta) - m_W^2 =0 \, .
    \end{equation}
Notice that this has the same form as in the complex scalar case in Eq.~\eqref{HJ-scalar}; therefore, we recover the Hawking spectrum. In this example, the non-minimal coupling to the electromagnetic field does not contribute. Moreover, the gauge condition ensures that the correct equation of motion is obtained, leading to the expected HR.

\subsection{Fermions} \label{Fermions-tunneling}
To describe the fermions of the SM, we use a KG-like equation given in Eq.~\eqref{KG-fermions12}. In this case, we use the standard WKB ansatz, which describes the spin-up or spin-down case, respectively:
    \begin{equation}
        \psi_{\uparrow} = \left[\ba{c} A\\
        0 \\
        B  \\
        0 \ea\right] \exp\left\{ \frac{i}{\hslash} I_{\uparrow} \right\} \, , \, \psi_{\downarrow} = \left[\ba{c} 0 \\C \\
        0 \\
        D  \ea\right] \exp\left\{ \frac{i}{\hslash} I_{\downarrow}  \right\} \, ,
    \end{equation}
where we omit the possible coordinate dependence of the functions $A$, $B$, $C$, and $D$.

For the spin-up case, we take the limit $\hslash \to 0$, then, for any coefficient $A$ or $B$, we obtain
    \begin{equation}
        g^{\mu\nu} (\partial_\mu I_{\uparrow} - qA_\mu)(\partial_\nu I_{\uparrow} - qA_\nu) - m_f^2 = 0 \, .
    \end{equation}
This equation is identical to the one obtained in Eq.~\eqref{Upsilon-q}, and therefore reproduces the usual HR. Notice that the same argument applies to the spin-down case as well. 

We emphasize that the key ingredient in our approach is the construction of a KG-like equation~\eqref{KG-fermions12}. In contrast, previous studies have solved the Dirac equation directly for rotating BHs~\cite{Jian2009fermions, Yale2011exact}. However, this procedure still requires a near-horizon approximation~\footnote{Some early results~\cite{Kerner2008fermions} showed that, for non-rotating BHs, separability is not a major complication due to the absence of mixing between the radial and polar angular coordinates.}, since the Dirac equation~\eqref{Motion.Eq.D} is not generically separable within the usual WKB ansatz. Instead, its separability relies on a \emph{Chandrasekhar ansatz} for the spinor solution~\cite{Chandrasekhar1976solution, Mukhopadhyay2000behaviour}. Moreover, by adopting a Chandrasekhar-like ansatz, some authors have demonstrated that the Hawking spectrum is recovered within this approximation~\cite{Deng2014hawking}. The geometric origin of this property is the hidden symmetry associated with the Killing--Yano tensor $Y_{\mu\nu}$, or equivalently the principal tensor of the Kerr geometry~\cite{Frolov2017black}. In particular, the corresponding second-rank Killing tensor is generated through the ``square'' of~\cite{Frolov2017black},
    \begin{equation}
        k_{\mu\nu} = Y_{\mu\alpha} {Y^\alpha}_\nu \, ,
    \end{equation}
which in turn gives rise to the Carter constant~\eqref{Carter-cte} and the separability of the HJ equation.
\section{Hawking spectrum for some theories beyond standard model} \label{sec: HR BSM}
In massive gravity, a massive graviton can be consistently described on a curved background. In this case, the field equations reduce, after fixing an appropriate gauge, to a KG–like equation~\eqref{KN-graviton}. To solve this equation, following the strategy used in previous examples, we adopt the WKB ansatz
    \begin{equation}
        h_{\mu\nu} = {\rm h}_{\mu\nu} \exp\left( \frac{i}{\hslash} I\right) \, ,
    \end{equation}
where ${\rm h}_{\mu\nu}={\rm h}_{\mu\nu}(t,\theta,\varphi,r)$, and suppress its explicit coordinate dependence in the following.

In the limit $\hslash \to 0$, we obtain the expression
     \begin{equation}
        [g^{\rho\sigma} (\partial_\rho I)(\partial_\sigma I) - m^2] {\rm h}_{\mu\nu}= 0 \, .
    \end{equation}
To ensure a physically consistent description, we impose the constraints in Eq.~\eqref{graviton-const} (see also Ref.~\cite{Buchbinder2000equations}). This leads to the nontrivial condition
    \begin{equation}
        \Upsilon(r,\theta) - m^2 =0 \, .
    \end{equation}
From this result, one recovers the correct Hawking spectrum. It is worth noting that, although the gauge condition is not explicitly required to reduce the equation of motion at this stage, it remains essential for consistently describing the five physical degrees of freedom of a massive graviton emitted by a KN black hole.

It is also important to note that the consistency of the description requires the presence of an additional term of the form
    \begin{equation}
        {f(R)_{\mu\nu}}^{\alpha\beta} = 2{{{R^\alpha}_\mu}^\beta}_\nu \, ,
    \end{equation}
where ${f(R)_{\mu\nu}}^{\alpha\beta}$ denotes a function of the Riemann tensor and its contractions. This contribution arises both from the non-commutativity of covariant derivatives and from possible non-minimal coupling terms, which are necessary to ensure causality and unitarity~\cite{Cuccieri1995tree}.

We now analyze the quantum tunneling of gravitinos with $q=0$. In this case, we consider the equation of motion~\eqref{KG-fermions32} and adopt the WKB ansatz   \begin{equation}
        \Psi_\lambda = \left[\ba{c} a_\lambda\\
        b_\lambda \\
        c_\lambda  \\
        d_\lambda \ea\right] \exp\left\{ \frac{i}{\hslash} I \right\} \, ,
    \end{equation}
where we have omitted the dependence on $t$, $\theta$, $\varphi$, and $r$ in the vector-spinor functions, $a_\lambda, b_\lambda, c_\lambda$ and $d_\lambda$.

Then, after some algebra, taking the limit $\hslash \to 0$, we obtain for each component
    \begin{equation}
        g^{\mu\nu} (\partial_\mu I)(\partial_\nu I) - m^2 = 0 \, .
    \end{equation}
As in all previous cases, it is straightforward to recover the Hawking spectrum. Notice that, in addition, a term involving the Riemann tensor appears, analogous to the graviton case; however, in the present case,
    \begin{equation}
        f(R) = - \frac{1}{8} \gamma^{\mu\nu} R_{\mu\nu\rho\sigma} \gamma^{\rho\sigma} \, .
    \end{equation}
It does not contribute to the Hawking spectrum and can be reduced to a term proportional to the Ricci scalar $R$.

In Ref.~\cite{Erbin2018universality}, following the prescription for free fields discussed in Subsec.~\ref{subsec: higher-spin} and based on Refs.~\cite{Singh1974lagrangian, Singh1974lagrangian2}, a generalized KG-like equation~\eqref{KG} is employed. However, this approach does not take into account the possible electric charge in the KN family of BHs. As we have seen in the case of charged $W$ bosons, the minimal coupling is necessary to reproduce the usual Hawking spectrum, and non-minimal couplings to the electromagnetic field are necessary to correctly describe the gyromagnetic moment, corresponding to a $g$-factor close to $g=2$~\cite{Ferrara1992g}.

Then, the generalization to higher spin fields $s>2$ will take the form:
    \begin{equation}
    \begin{aligned}
        &g^{\mu\nu} D_\mu D_\nu \Phi_{\mu_1 \cdots \mu_s} + {f(R)_{\mu_1 \cdots \mu_s}}^{\nu_1 \cdots \nu_s} \Phi_{\nu_1 \cdots \nu_s} \\
        &+ {j(F)_{\mu_1 \cdots \mu_s}}^{\nu_1 \cdots \nu_s} \Phi_{\nu_1 \cdots \nu_s} + \frac{M}{\hslash^2} \Phi_{\mu_1 \cdots \mu_s} = 0 \, ,
    \end{aligned}
    \end{equation}
where $D_\mu = \nabla_\mu - iqA_\mu/\hslash$, with  $f(R)$ a function of the Riemann tensor and its contractions, arising both from anticommutation of covariant derivatives and from non-minimal coupling terms. Similarly, $j(F)$ is a function of the electromagnetic tensor related to non-minimal coupling.

In principle, the function $j(F)$ is unknown. Since several complications already arise in the simple case of a spin $s=3/2$ field with electromagnetic interactions in flat spacetime~\cite{Porrati2009causal}, similar inconsistencies are expected to occur even when the constraints needed to eliminate the auxiliary fields are imposed~\cite{Erbin2018universality}.
    \begin{equation}
        \nabla^{\mu_1} \Phi_{\mu_1 \cdots \mu_s} = 0 \quad \text{and} \quad g^{\mu_1\mu_2} \Phi_{\mu_1 \mu_2 \cdots \mu_s} = 0 \, .
    \end{equation}
To date, several proposals have been put forward to generalize higher-spin fields in the presence of electromagnetic interactions, although they are typically restricted to constant electromagnetic backgrounds~\cite{Benakli2023spin, Delplanque2024massive, Benakli2026effective}. Consequently, constructing a fully consistent theory remains a challenging problem. As in our previous examples, we recover the correct Hawking spectrum; however, ensuring the consistent propagation of the physical degrees of freedom remains an open issue.
\section{Benchmarking and Outlook} \label{sec: HR-AC}
First, we present a direct comparison with related works. Instead of working directly with the metric~\eqref{KN-metric}, one can consider the change of variable $\chi = \varphi - \Omega^\prime t$, where $\Omega^\prime = [a (2Mr - Q^2)]/[(r^2 + a^2)^2 - \Delta a^2 \sin^2\theta]$ is the angular velocity of the BH. The KN spacetime contains a timelike ring singularity hidden behind the event horizon. As an observer approaches the horizon, time appears to stop from the perspective of a distant observer, and signals originating near the horizon become increasingly redshifted. Then $\Omega_H = \Omega^\prime \, |_{r = r_H}$ determines how the particle's angular momentum affects the measurement of its angular coordinate $\varphi$ by an observer at infinity, taking into account the BH's rotation. This term effectively redefines the coordinates, allowing the behavior of particles near the horizon, where relativistic effects such as gravitational redshift and frame dragging are significant, to be analyzed consistently.

From the previous statement, in Refs.~\cite{Jian2009fermions, Li2015massive}, a change of variables is used to determine the imaginary part of $I$ in a \emph{dragging coordinate system}. However, we would like to remark that the fixed polar angle approximation is still employed. In our case, we use ${\cal R}(r)$, which depends explicitly only on $r$. In contrast, in Refs.~\cite{Jian2009fermions, Li2015massive}, the function $W(r)$ is employed; nevertheless, these function also depends on the variable $\theta$ variable, although it is not specified whether $\theta$ is fixed, in fact, in that prescription the $\theta$-dependence vanishes only at the event horizon $r=r_H$. By contrast, in our approach, the separation of variables arises in a completely natural way due to the existence of the Carter constant ${\cal K}$.

Another approach suggests that the limit $\hslash \to 0$ is not necessary; instead, it is argued that the condition $g^{tt} \to 0$ is sufficient, and for rotating BHs, this procedure works only after redefining the energy-related coordinate through $\chi$ and fixing $\theta=\theta_0$~\cite{Yale2011exact}. Then, implicitly assumes that only the $t-r$ sector contributes near the event horizon. In contrast, the treatment presented here reveals that, once the hidden symmetry of the KN spacetime (and its particular limits, such as the Kerr BH) is properly taken into account, the $t-r$ contribution to the imaginary part of the semiclassical action $I$ emerges naturally as a consequence of the underlying spacetime structure. Therefore, the complex-path approach naturally reduces the background to an effective $(1+1)$-dimensional blackbody at the event horizon. This is the so-called \emph{dimensional reduction} underlying the derivation of Hawking radiation via anomaly cancellation~\cite{Robinson2005relationship, Iso2006anomalies, Iso2006hawking}, where mass and interaction terms in the action are neglected, as the kinetic term is expected to dominate in the high-energy regime close to the event horizon~\cite{Umetsu2010hawking}. 

A particularly interesting aspect of this result is that the dimensional reduction approach relies on the separability of the KG-like equation used in the action functional~\cite{Iso2006hawking, Umetsu2010hawking}. However, several spacetimes extending the Kerr geometry preserve the hidden symmetries and, consequently, the Carter constant, thereby retaining the separability of the HJ equation~\cite{Papadopoulos2018preserving, Carson2020asymptotically, Papadopoulos2021kerr, Junior2021can}. This observation allows us to relax the requirement of KG separability, since separability of the HJ equation does not necessarily imply separability of the KG equation~\cite{Papadopoulos2021kerr}.

The method developed here can be extended to a broad class of BHs. In particular, the Plebański–Demiański family~\cite{Plebanski1976rotating} demonstrates that many solutions beyond Kerr, including Kerr–Newman–NUT–(A)dS black holes and several string-inspired geometries, possess the same separability structure~\cite{Griffiths2006new, Podolsky2006accelerating}, enabling the introduction of a Carter-type constant and the implementation of HJ analyses analogous to those in the Kerr spacetime. 

The accelerating sector~\cite{Griffiths2005accelerating}, however, requires a more careful treatment, as the hidden symmetry underlying separability may differ from that of the non-accelerating geometries. When $\alpha$ denotes the acceleration parameter, the Hawking temperature obtained from the surface gravity has appeared in the literature with different overall factors~\cite{Gillani2011hawking, Bilal2013thermodynamics, Vanzo2011tunn}. In several tunneling calculations for accelerating black holes~\cite{Gillani2011tunneling,Rehman2011charged,Lin2015accelerating}, the result is given by
    \begin{equation}
        T_H=\frac{(r_{+}-M)}{2\pi(r_{+}^{2}+a^{2})}\left(1-\alpha^{2}r_{+}^{2}\right) \, . \label{T_H_rotating}
    \end{equation}
This expression agrees with more recent thermodynamic analyses~\cite{Appels2016thermodynamics, Anabalon2019thermodynamics}, while the framework developed here may help clarify the derivation of the spectrum. We hope that the result given in Eq.~\eqref{T_H_rotating} corresponds to the correct Hawking temperature.

\section{Conclusions} \label{sec: Conclusion}
In this paper, we have presented a systematic derivation of Hawking radiation as a quantum tunneling process across the event horizon of the KN family of black holes. We have shown that near-horizon approximations involving the polar angle are unnecessary, since the HJ equations remain separable and can be treated exactly. 

The main result of this work underscores the central role of symmetries in both field theory and spacetime geometry. When a well-defined wave equation is available ( e.g., via gauge symmetries), the correspondence principle reduces the problem to quantum tunneling using the complex path method. Equally important are the hidden geometric symmetries underlying HJ equation separability, rather than approaches based on dimensional reduction induced by the separability of the KG equation in the curved-spacetime action functional. As a consequence, the effective reduction to a $(1+1)$-dimensional blackbody description near the horizon emerges naturally. The particle mass $m_t$ does not affect the Hawking spectrum, although it still appears as a threshold condition, $\omega \geq m_t$.

For fermionic fields, the absence of a rank-two Killing tensor associated directly with the Dirac equation complicates the semiclassical analysis. Nevertheless, by exploiting the Killing--Yano tensor, we have shown that the dynamics can be reformulated in terms of a KG-like equation, allowing the standard Hawking spectrum to be recovered in a transparent manner.

The extension of these methods to higher-spin theories remains considerably more challenging. In particular, a consistent description requires propagating the correct physical degrees of freedom for charged fields interacting with external electromagnetic backgrounds. Establishing such a framework is a necessary step toward a complete understanding of HR for arbitrary-spin particles in rotating black-hole spacetimes.

\begin{acknowledgments}
I.P.C. thanks SECIHTI for the scholarship as ``Ayudante de Investigador SNII III'' and thanks Hugo García Compeán for the earlier discussion at the beginning of this work.
\end{acknowledgments}

\clearpage
\onecolumngrid

\appendix

\section{On the separability of Hamilton-Jacobi equation} \label{AppA}
To show the separability of the HJ equation, we consider a simple case of a complex scalar field, where the action is given by
    \begin{equation}
        {\cal S}_{\Phi} = \frac{1}{2} \int d^4 x \sqrt{-g} \left(g^{\mu\nu} D_\mu \Phi D_\nu \Phi^{*} - \frac{m_{\Phi}^2}{\hslash^2} |\Phi|^2\right) \, , \label{S-scalar}
    \end{equation}
where $D_\mu = \nabla_\mu - iq A_\mu/ \hslash$ is the covariant gauge derivative. 

From Eq.~\eqref{S-scalar}, we can derive the Klein-Gordon equation~\eqref{KG}. Notice that the Eq.~\eqref{KG}, after to using the ansatz~\eqref{s-action-ansatz}, can be written as
    \begin{equation}
        0 = g^{tt}(\omega + qA_t )^2 - 2g^{t\varphi}(\omega + qA_t ) (\ell - qA_\varphi) - m_\Phi^2 + g^{\theta\theta}(\partial_\theta W)^2 + g^{\varphi\varphi}(\ell - qA_\varphi)^2 + g^{rr}(\partial_r W)^2 \, .
    \end{equation}
We focus on the mixing part with the electromagnetic elements,
we have
    \begin{equation}
        L = g^{tt}(\omega + qA_t )^2 - 2g^{t\varphi}(\omega + qA_t ) (\ell - qA_\varphi) + g^{\varphi\varphi}(\ell - qA_\varphi)^2 \, .
    \end{equation}
Multiplying by $\Sigma(r,\theta)$, and defining $\tilde{\omega}= (\omega + qA_t)$ and $\tilde{\ell}= (\ell - qA_\varphi)$, from the metric given in the BL coordinates presented in Eq.~\eqref{KN-metric}-\eqref{KN-LE}, we obtain
    \begin{equation}
        \Sigma(r,\theta) L = \frac{[(r^2 + a^2) \tilde{\omega}- a\tilde{\ell}]^2}{\Lambda} - \left[ a\sin\theta \tilde{\omega} - \frac{\tilde{\ell}}{\sin\theta} \right]^2 \, ,
    \end{equation}
where we have used
    \begin{equation*}
    \begin{aligned}
        \frac{\Sigma(r,\theta)}{F(r,\theta)} &= \frac{[(r^2 + a^2)^2 - a^2 \Delta \sin^2 \theta]}{\Delta} = \frac{(r^2 + a^2)^2}{\Delta} - a^2 \sin^2 \theta \, ,\\ 
        \frac{H(r,\theta)}{K(r,\theta)} &= \frac{a [(r^2 + a^2) - \Delta] }{[(r^2 + a^2)^2 - a^2\Delta \sin^2 \theta]} \Rightarrow \frac{H(r,\theta) \Sigma(r,\theta)}{K(r,\theta)F(r,\theta)} = \frac{a [(r^2 + a^2) - \Delta]}{\Delta} \, , \\
        \frac{f(r,\theta)}{K(r,\theta)} &= \frac{\Delta-a^2\sin^2\theta}{\sin^2 \theta[(r^2 + a^2)^2 - a^2 \Delta \sin^2 \theta]} \Rightarrow \frac{f(r,\theta)\Sigma(r,\theta)}{K(r,\theta)F(r,\theta)} = \frac{\Delta-a^2\sin^2\theta}{\sin^2 \theta \Delta} = \frac{1}{\sin^2\theta} - \frac{a^2}{\Delta} \, ,\\
        \Sigma(r, \theta) G(r,\theta) &= \Delta(r) = \Delta \, .
    \end{aligned}
    \end{equation*}
Notice now that we have
    \begin{equation}
        (r^2 + a^2) \tilde{\omega}- a\tilde{\ell} = (r^2 + a^2) \omega - a\ell - q Qr \quad \text{and} \quad a\sin\theta \tilde{\omega} - \frac{\tilde{\ell}}{\sin\theta} = a\sin\theta \omega - \frac{\ell}{\sin\theta} \, .
    \end{equation}
Then, the HJ equation could be written as (remember that $\Delta=\Delta(r) = (r-r_{-})(r-r_{+})$),
    \begin{equation}
        0 = \frac{[(r^2 + a^2) \omega - a\ell - q Qr]^2}{\Delta} - \left[ a\sin\theta \omega - \frac{\ell}{\sin\theta} \right]^2 - (\partial_\theta W)^2 - \Delta (\partial_r W)^2 - (r^2 +a^2 \cos^2\theta) m_\Phi^2 \, .
    \end{equation}
Therefore, now es evident the HJ equation is separable, and we can use $W(r,\theta) = {\cal R}(r) + {\cal T}(\theta)$ to write Eq.~\eqref{HJ-separable}.

Notice that the result is valid for $q=0$; in that case, we have defined the $\Upsilon(r,\theta)$ expression in Eq.~\eqref{Upsilon}.

On the other hand, in classical mechanics, we can define
    \begin{equation}
        \frac{d I}{d\theta} = \frac{d{\cal T}}{d\theta} = p_\theta \, .
    \end{equation}
Then, the constant of separation of variables $\lambda$ can be written as
    \begin{equation}
        \lambda = p_\theta^2 + \cos^2\theta \left[ a^2(m_\Phi^2-\omega^2) + \left( \frac{\ell}{\sin\theta} \right)^2\right] + (a\omega - \ell)^2 \, .
    \end{equation}
From Carter's original derivation of the Kerr BH~\cite{Carter1968global}, which can be generalized to the KN family~\cite{Carter2009republication}, the constant of separation of variables can be determined by the Stackel-Killing tensor $K_{\mu\nu}$ as Eq.~\eqref{Carter-cte}, that have the form ${\cal K} = K- 2a\omega \ell$
    \begin{equation}
        K = p_\theta^2 + \left( \frac{\ell}{\sin\theta} \right)^2 + a^2 m_\Phi^2 \cos^2\theta + a^2 \omega^2 \sin^2\theta \, . \label{K}
    \end{equation}
After a little algebra is easy to show that $\lambda = K- 2a\omega \ell$.

To obtain a physical interpretation, one may redefine the constant $\Lambda = K + m_\Phi^2 a^2$, which can be understood as the algebraic sum of the squared contributions to the particle’s total angular momentum. These contributions arise not only from the angular motion, but also from the radial motion~\cite{De1999meaning}.

\section{Back-reaction and Bekenstein Hawking entropy} \label{AppB}
The well-known Bekenstein-Hawking entropy area law for the entropy~\cite{Bekenstein1973black, Hawking1975particle} of a black hole is $S=A/4$, where $A$ is the area of the event horizon. In the case of the KN family, $A= 4\pi (r_H^2 + a^2)$ with $r_H=r_{+}$ given in Eq.~\eqref{Horizonts}.

If we consider the back-reaction due to the emission of a particle by a KN black hole, the imaginary part of the semiclassical action becomes
    \begin{equation}
        2Im I = \int_\gamma \frac{d }{dE} \left(\frac{E}{T_H} \right)dE = \int_{(0,0,0)}^{(\omega,\ell, q)} \frac{d }{dE} \left(\frac{E}{T_H} \right) (d\omega^\prime  - \Omega_H d\ell^\prime + \Phi_H dq^\prime) \, ,
    \end{equation}
where $\gamma$ describes the path of integration from $(0,0,0)$ to $(\omega,\ell, q)$. Then, we can rewrite the imaginary part as
    \begin{equation}
    \begin{aligned}
        2Im I &= \int_{(0,0,0)}^{(\omega,\ell, q)} \frac{2\pi [(M + \sqrt{M^2 - a^2 - Q^2})^2 + a^2]}{\sqrt{M^2 - a^2 - Q^2}} (d\omega^\prime  - \Omega_H d\ell^\prime + \Phi_H dq^\prime) \\
        &= -2\pi \int_{(M,J,q)}^{(M-\omega, J-\ell , Q-q)} \frac{(r_H^{\prime 2} + a^{\prime 2})}{(r_H^\prime - M^\prime)} \bigg[dM^\prime - \frac{a^\prime}{(r_H^{\prime 2} + a^{\prime 2})} dJ^\prime - \frac{Q^\prime r_H^\prime}{(r_H^{\prime 2} + a^{\prime 2})} d Q^\prime \bigg] \, .
    \end{aligned}
    \end{equation}
Here, we have used the change of variables $M^\prime = M-\omega^\prime$, $Q^\prime = Q- q^\prime$, $J^\prime = J - \ell^\prime$, which defines
    \begin{equation}
        a^\prime = \frac{J^\prime}{M^\prime} \quad \text{and} \quad r_H^\prime = M^\prime + \sqrt{M^{\prime 2} - a^{\prime 2} - Q^{\prime 2}} \, .
    \end{equation}
From this, we obtain the following differentials
    \begin{equation}
         da^\prime = \frac{dJ^\prime}{M^\prime} - \frac{a^\prime}{M^\prime} dM^\prime \quad \text{and} \quad dr_H^\prime = dM^\prime + \frac{M^\prime dM^\prime - a^\prime da^\prime - Q^\prime dQ^\prime }{(r_H^\prime - M^\prime)} \, .
    \end{equation}
Then, we can show that
    \begin{equation}
        \frac{(r_H^{\prime 2} + a^{\prime 2})}{(r_H^\prime - M^\prime)} \bigg[dM^\prime - \frac{a^\prime}{(r_H^{\prime 2} + a^{\prime 2})} dJ^\prime - \frac{Q^\prime r_H^\prime}{(r_H^{\prime 2} + a^{\prime 2})} d Q^\prime \bigg] = \frac{1}{2} d(r_H^{\prime 2} + a^{\prime 2}) \, .
    \end{equation}
Therefore,
    \begin{equation}
        2Im I = -\pi \int_{(M,J,q)}^{(M-\omega, J-\ell , Q-q)} d(r_H^{\prime 2} + a^{\prime 2}) = - \Delta S_{B-H} = - [S(M-\omega, J-\ell , Q-q) - S(M, J, Q)]\, .
    \end{equation}
where the Bekenstein-Hawking entropy is given in Eq.~\eqref{BH-entropy}.
\section{Charged \texorpdfstring{$W$}{W} bosons} \label{AppC}
For $W^{\pm}$ bosons, we can generalize to curved spacetime through the action that considers how $\hslash$ appears:
    \begin{equation}
        {\cal S}_{W} = \int dx^4 \sqrt{-g} \left( - \frac{1}{2} F_{\mu\nu}^{+} F^{- \mu\nu} + \frac{m_W^2}{\hslash^2} W^+_\mu W^{-\mu} - i \frac{q}{\hslash} F_{\mu\nu} W^{+\mu} W^{-\nu} \right) \, ,
    \end{equation}
where $F_{\mu\nu}^{\pm} = {\cal D}_\mu^{\pm} W^{\pm}_\nu - {\cal D}_\nu^{\pm} W^{\pm}_\mu$, with $\nabla_\mu W^{\pm}_\nu = \partial_\mu^{\pm} W^{\pm}_\nu - \Gamma_{\mu\nu}^\lambda W^{\pm}_\lambda$, with ${\cal D}_\mu^{\pm} = \nabla_\mu \pm ie A_\mu/\hslash$. We obtain the following motion equation,
    \begin{equation}
        {\cal D}_\mu^{\pm} F^{\pm \mu\nu} + \frac{m_W^2}{\hslash^2} W^{\pm \nu} \pm \frac{i}{\hslash} e F^{\nu\mu} W_\mu^{\pm} =0 \, , \label{W-ME}
    \end{equation}
For the $W$ boson with charge $+e$, we can rewrite the motion equation~\eqref{W-ME}  as
    \begin{equation}
        \nabla_\mu F^{+ \mu\nu} + \frac{m_W^2}{\hslash^2} W^{+ \nu} \pm \frac{i}{\hslash} eA_\mu F^{+ \mu\nu} + \frac{i}{\hslash} e F^{\nu\mu} W_\mu^{+} =0 \, .
    \end{equation}
From the solution proposed in Eq.~\eqref{WKB-W}, as $F^{+\mu\nu}$ is antisymmetric, we can write
    \begin{equation}
    \begin{aligned}
        \nabla_\mu F^{+ \mu\nu} &= \frac{1}{\sqrt{-g}} \partial_\mu \left(\sqrt{-g}F^{+ \mu\nu} \right) =  \frac{1}{\sqrt{-g}} \partial_\mu \left[\sqrt{-g} g^{\mu\rho} g^{\nu\lambda} \left(\nabla_\rho W_\lambda^{+} + \frac{i}{\hslash} e A_\rho W_\lambda^{+}  - \nabla_\lambda W_\rho^{+} - \frac{i}{\hslash} eA_\lambda W_\rho^{+} \right)\right] \\
        &= \frac{1}{\sqrt{-g}} [\partial_\mu (\sqrt{-g} g^{\mu\rho} g^{\nu\lambda})] (\partial_\rho W_\lambda^{+} - \partial_\lambda W_\rho^{+}) +   (g^{\mu\rho} g^{\nu\lambda}) \partial_\mu (\partial_\rho W_\lambda^{+} - \partial_\lambda W_\rho^{+}) \\
        &\quad  + \frac{i}{\hslash} e \frac{1}{\sqrt{-g}} \left[\partial_\mu (\sqrt{-g} g^{\mu\rho} g^{\nu\lambda}) \right] \left( A_\rho W_\lambda^{+} - A_\lambda W_\rho^{+} \right)  + \frac{i}{\hslash} e (g^{\mu\rho} g^{\nu\lambda}) \partial_\mu \left( A_\rho W_\lambda^{+} - A_\lambda W_\rho^{+} \right) \, .
    \end{aligned}
    \end{equation}
and
    \begin{equation}
        \frac{i}{\hslash} e A_\mu F^{+ \mu \nu} = \frac{i}{\hslash} e A_\mu g^{\mu\rho} g^{\nu\lambda} F_{\rho\lambda}^{+} = \frac{i}{\hslash} e A_\mu g^{\mu\rho} g^{\nu\lambda} \left( \partial_\rho W_\lambda^{+} - \partial_\lambda W_\rho^{+}\right) -\frac{e^2}{\hslash^2} A_\mu g^{\mu\rho} g^{\nu\lambda} \left( A_\rho W_\lambda^{+} - A_\lambda W_\rho^{+}\right) \, .
    \end{equation}
Now, for the limit $\hslash \to 0$, one gets
    \begin{equation}
    \begin{aligned}
        0&= -(g^{\mu\rho} g^{\nu\lambda}) (\partial_\mu I) (w_{\lambda}^{+} \partial_\rho I - w_{\rho}^{+} \partial_\lambda I) - e (g^{\mu\rho} g^{\nu\lambda}) \left( A_\rho w_{\lambda}^{+} - A_\lambda w_{\rho}^{+} \right) (\partial_\mu I) \\
        &\quad  - e A_\mu g^{\mu\rho} g^{\nu\lambda} \left( w_{\lambda}^{+} \partial_\rho I - w_{\rho}^{+} \partial_\lambda I  \right) - e^2 A_\mu g^{\mu\rho} g^{\nu\lambda} \left( A_\rho w_{\lambda}^{+} - A_\lambda w_{\rho}^{+}\right) + m_W^2 g^{\nu\lambda}w_{\lambda}^{+} \\
        &= -(g^{\mu\rho} g^{\nu\lambda}) (\partial_\mu I + eA_\mu) \left[ \left( \partial_\rho I + eA_\rho \right) w_{\lambda}^{+} -  e\left( \partial_\lambda I + eA_\lambda \right)w_{\rho}^{+} \right] + m_W^2 g^{\nu\lambda}w_{\lambda}^{+} \, . \label{W-Eqs}
    \end{aligned}
    \end{equation}
Then, we obtain the expression presented in Eq.~\eqref{W-non-gauge}.

\twocolumngrid
\bibliography{References}

\end{document}